\documentclass[english, 11pt]{article}
\pdfoutput=1
\usepackage[utf8]{inputenc}
\usepackage[T1]{fontenc}
\usepackage{babel}
\usepackage{helvet}

\usepackage{graphicx}
\usepackage{fancyhdr}
\usepackage{tabularx}
\usepackage{multicol}
\usepackage{booktabs}
\usepackage{colortbl}
\usepackage[font=footnotesize, labelfont=bf, singlelinecheck=false, labelsep=period, skip=6pt]{caption}
\usepackage{xcolor}
\usepackage{hyperref}
\definecolor{urlcolor}{rgb}{0.11372549, 0.470588235, 0.11372549}
\definecolor{linkcolor}{rgb}{0.117647059, 0.564705882, 1}
\definecolor{citecolor}{rgb}{1, 0.270588235, 0}
\definecolor{grey}{rgb}{0.5, 0.5, 0.5}
\definecolor{lightblue}{rgb}{0.7, 0.7, 0.7}
\definecolor{lightergrey}{rgb}{0.9, 0.9, 0.9}
\hypersetup{breaklinks, colorlinks=true, citecolor=citecolor, linkcolor=linkcolor, urlcolor=urlcolor}

\addto\captionsenglish{}
\usepackage[superscript]{cite}
\usepackage[width=8.5in, height=11in, margin=1in]{geometry}
\newlength{\intercolsep}
\usepackage[compact,explicit]{titlesec}
\titleformat{\section}[block]{\bfseries\Large}{\thesection.}{1ex}{#1}[]
\titlespacing*{\section}{0pt}{*4.0}{*0.0}
\titleformat{\subsection}[block]{\bfseries\large}{\thesubsection.}{1ex}{#1}[]
\titlespacing*{\subsection}{0pt}{*2.0}{*-1.5}
\titleformat{\subsubsection}[block]{\bfseries}{\thesubsubsection.}{1ex}{#1}[]
\titlespacing*{\subsubsection}{0pt}{*1.0}{*-1.5}
\titleformat{\paragraph}[runin]{\bf}{}{0em}{#1.}[]
\titlespacing*{\paragraph}{0pt}{*0.0}{1ex}

\makeatletter \renewcommand\@biblabel[1]{#1.} \makeatother
\usepackage{changepage}
\usepackage{setspace}
\newcommand{\ObjTables}{\textit{ObjTables}}

\begin{document}


\title{
    \vspace{-2.4em}
    {\ObjTables}: structured spreadsheets that promote data quality, reuse, and integration
    \vspace{-1em}
}

\author{
    Jonathan R. Karr\textsuperscript{1,2{*}}, 
    Wolfram Liebermeister\textsuperscript{3},
    Arthur P. Goldberg\textsuperscript{1,2},\\
    John A. P. Sekar\textsuperscript{1,2} \&
    Bilal Shaikh\textsuperscript{1,2}
    \vspace{-1em}
}

\date{
    August 6, 2020
    \vspace{-1em}
}

\maketitle

\begin{adjustwidth}{0.75em}{2.5em}
\setlength{\parskip}{0.5em}

\begin{itemize}
\setlength{\labelsep}{0pt}
\setlength{\labelwidth}{0pt}
\setlength{\itemsep}{4pt}
\item[\textsuperscript{1}]Icahn Institute for Data Science and Genomic Technology, Icahn School of Medicine at Mount Sinai, New York, NY 10029, USA.

\item[\textsuperscript{2}]Department of Genetics and Genomic Sciences, Icahn School of Medicine at Mount Sinai, New York, NY 10029, USA.

\item[\textsuperscript{3}]Universit\'e Paris-Saclay, INRAE, MaIAGE, 78350, Jouy-en-Josas, France.

\item[{*}]Corresponding author: Jonathan Karr (\href{mailto:karr@mssm.edu}{karr@mssm.edu})
\end{itemize}
\end{adjustwidth}



Many scientific problems require integrating multiple types and sources of data \cite{gurevitch2018meta}. As one of the most common media for scientific data, supplementary spreadsheets associated with articles are a key resource \cite{greenbaum2017structuring}. Spreadsheets are popular because they are easy for authors to write with software such as Excel, LibreOffice, and Google Sheets and easy for journals to archive.

However, spreadsheets are often difficult for other investigators to reuse \cite{pop2015use, greenbaum2017structuring, cheifet2018making}. First, spreadsheets frequently capture information ad hoc because spreadsheets only support a few data types, and there are no universal conventions for encoding multi-dimensional data and metadata into spreadsheets. Second, the ad hoc nature of spreadsheets fosters errors \cite{powell2008critical, ziemann2016gene}. Due to their popularity, the limited reusability of spreadsheets hinders a wide range of research. For example, this inhibits meta-analyses of multiple studies, comparative analyses of multiple organisms, and integrative research such as whole-cell modeling \cite{goldberg2018emerging}.

To facilitate comparative and integrative research, we encourage authors to write their data and metadata into spreadsheets systematically and thoroughly check their spreadsheets for errors. Several fields of science have begun to develop standards for rigorously encoding domain-specific data into spreadsheets such as ISA-Tab for experimental studies \cite{sansone2008first} (\textcolor{linkcolor}{Section~4 of the Supplementary Information}). However, it remains challenging to reuse most scientific spreadsheets.

\begin{figure}[!p]
    \centerline{\includegraphics{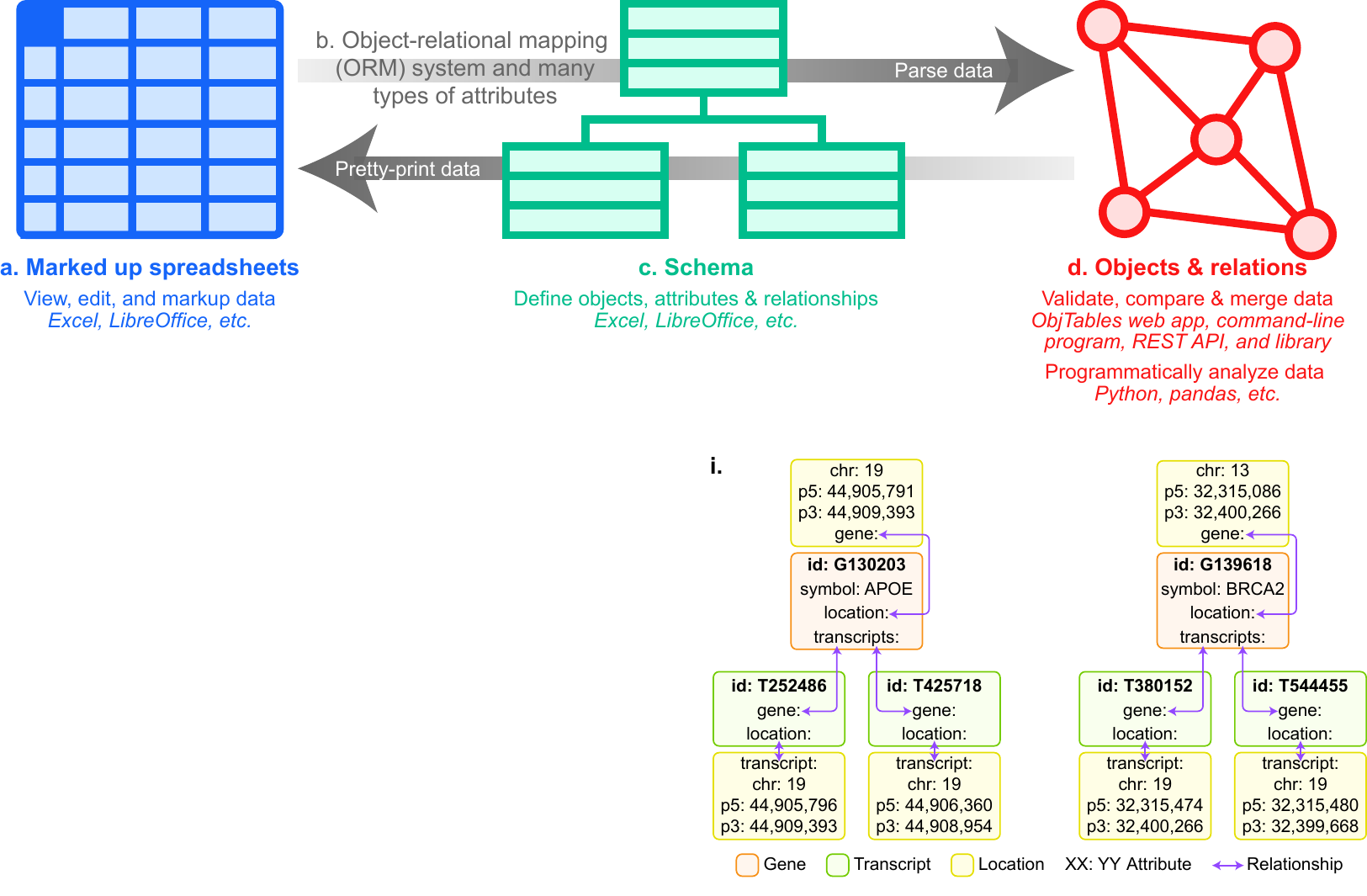}}
    
    \vspace{-2.4096in}
    
    \tiny
    \aboverulesep = 0mm
    \belowrulesep = 0mm
    \setlength{\tabcolsep}{2.0pt}
    
    \textbf{\scriptsize e. `!!\_Table of contents' worksheet}\\[0.1em]
    \setlength{\intercolsep}{6.75pt}
    \begin{tabularx}{\glueexpr0.5\textwidth - 9.75px}{
    @{}|
    l
    !{\color{lightgray}\vrule}l
    !{\color{lightgray}\vrule}l
    |@{}}
    \toprule
    \multicolumn{3}{@{}|l|@{}}{\color{grey}!!!ObjTables objTablesVersion=`1.0.0' author=`John Doe' date=`2020-05-01'} \\
    \arrayrulecolor{lightgray}\midrule
    \arrayrulecolor{black}
    \multicolumn{3}{@{}|l|@{}}{\color{grey}!!ObjTables type=`TableOfContents'} \\
    \arrayrulecolor{black}\midrule
    \multicolumn{1}{|l}{\cellcolor{lightblue}{\bf !Worksheet\hspace{\intercolsep}\hphantom{}}} & 
    \multicolumn{1}{l}{\cellcolor{lightblue}{\bf !Description\hspace{\intercolsep}\hphantom{}}} & 
    \multicolumn{1}{l|}{\cellcolor{lightblue}{\bf !Objects\hspace{\intercolsep}\hphantom{}}} \\
    \arrayrulecolor{black}\midrule
    Genes 
    \hspace{\intercolsep} & Genes in the genome 
    \hspace{\intercolsep} & 2 
    \hspace{\intercolsep} \\
    \arrayrulecolor{lightgray}\midrule
    
    Transcript variants 
    \hspace{\intercolsep} & Splice variants expressed from the genome 
    \hspace{\intercolsep} & 4 
    \hspace{\intercolsep} \\
    \arrayrulecolor{black}\bottomrule
    \end{tabularx}

    \vspace{1em}

    \textbf{\scriptsize f. `!!Genes' worksheet}\\[0.1em]
    \setlength{\intercolsep}{4.1pt}
    \begin{tabularx}{\glueexpr0.5\textwidth - 9.75px}{
    @{}|
    l!{\color{lightgray}\vrule}
    l!{\color{lightgray}\vrule}
    l!{\color{lightgray}\vrule}
    l!{\color{lightgray}\vrule}
    l
    |@{}}
    \toprule
    \multicolumn{5}{@{}|l|@{}}{\color{grey}!!ObjTables type=`Data' class=`Gene'} \\
    \midrule
    & & \multicolumn{3}{c|}{\cellcolor{lightblue}{\bf !Location}} \\ \cmidrule{3-5}
    \multicolumn{1}{|l}{\cellcolor{lightblue}{\bf !Id\hspace{\intercolsep}\hphantom{}}} 
    & \multicolumn{1}{l}{\cellcolor{lightblue}{\bf !Symbol\hspace{\intercolsep}\hphantom{}}}
    & \multicolumn{1}{l}{\cellcolor{lightblue}{\bf !Chromosome\hspace{\intercolsep}\hphantom{}}} 
    & \multicolumn{1}{l}{\cellcolor{lightblue}{\bf !5'\hspace{\intercolsep}\hphantom{}}} 
    & \multicolumn{1}{l|}{\cellcolor{lightblue}{\bf !3'\hspace{\intercolsep}\hphantom{}}} \\
    \midrule
    ENSG00000130203
    \hspace{\intercolsep} & APOE 
    \hspace{\intercolsep} & 19 
    \hspace{\intercolsep} & 44,905,791 
    \hspace{\intercolsep} & 44,909,393 
    \hspace{\intercolsep} \\
    
    \arrayrulecolor{lightgray}\midrule
    
    ENSG00000139618 
    \hspace{\intercolsep} & BRCA2 
    \hspace{\intercolsep} & 13
    \hspace{\intercolsep} & 32,315,086
    \hspace{\intercolsep} & 32,400,266 
    \hspace{\intercolsep} \\
    \arrayrulecolor{black}\bottomrule
    \end{tabularx}
    
    \vspace{1em}

    \textbf{\scriptsize g. `!!Transcript variants' worksheet}\\[0.1em]
    \setlength{\intercolsep}{0.9pt}
    \begin{tabularx}{\glueexpr0.5\textwidth - 9.75px}{
    @{}|
    l
    !{\color{lightgray}\vrule}l
    !{\color{lightgray}\vrule}l
    !{\color{lightgray}\vrule}l
    !{\color{lightgray}\vrule}l
    |@{}}
    \toprule
    \multicolumn{5}{@{}|l|@{}}{\color{grey}!!ObjTables type=`Data' class=`Transcript'} \\
    \midrule
    & & \multicolumn{3}{c|}{\cellcolor{lightblue}{\bf !Location}} \\ \cmidrule{3-5}
    \multicolumn{1}{|l}{\cellcolor{lightblue}{\bf !Id\hspace{\intercolsep}\hphantom{}}} & 
    \multicolumn{1}{l}{\cellcolor{lightblue}{\bf !Gene\hspace{\intercolsep}\hphantom{}}} & 
    \multicolumn{1}{l}{\cellcolor{lightblue}{\bf !Chr\ldots\hspace{\intercolsep}\hphantom{}}} & 
    \multicolumn{1}{l}{\cellcolor{lightblue}{\bf !5'\hspace{\intercolsep}\hphantom{}}} & 
    \multicolumn{1}{l|}{\cellcolor{lightblue}{\bf !3'\hspace{\intercolsep}\hphantom{}}} \\
    \midrule
    
    ENST00000252486.9 
    \hspace{\intercolsep} & ENSG00000130203 
    \hspace{\intercolsep} & 19 
    \hspace{\intercolsep} & 44,905,796 
    \hspace{\intercolsep} & 44,909,393 
    \hspace{\intercolsep} \\
    \arrayrulecolor{lightgray}\midrule
    
    ENST00000425718.1 
    \hspace{\intercolsep} & ENSG00000130203 
    \hspace{\intercolsep} & 19 
    \hspace{\intercolsep} & 44,906,360 
    \hspace{\intercolsep} & 44,908,954 
    \hspace{\intercolsep} \\
    \midrule
    
    ENST00000380152.7 
    \hspace{\intercolsep} & ENSG00000139618 
    \hspace{\intercolsep} & 13 
    \hspace{\intercolsep} & 32,315,474 
    \hspace{\intercolsep} & 32,400,266 
    \hspace{\intercolsep} \\
    \midrule
    
    ENST00000544455.5 
    \hspace{\intercolsep} & ENSG00000139618 
    \hspace{\intercolsep} & 13 
    \hspace{\intercolsep} & 32,315,480 
    \hspace{\intercolsep} & 32,399,668 
    \hspace{\intercolsep} \\
    \arrayrulecolor{black}\bottomrule
    \end{tabularx}
    
    \vspace{1em}

    \setlength{\intercolsep}{36.5pt}
    \textbf{\scriptsize h. `!!\_Schema' worksheet}\\[0.1em]
    \begin{tabularx}{\textwidth}{
    @{}|
    l
    !{\color{lightgray}\vrule}l
    !{\color{lightgray}\vrule}l
    !{\color{lightgray}\vrule}l
    !{\color{lightgray}\vrule}l
    |@{}}
    \toprule
    \multicolumn{5}{@{}|l|@{}}{\color{grey}!!ObjTables type=`Schema'} \\
    \midrule
    \multicolumn{1}{|l}{\cellcolor{lightblue}{\bf !Name\hspace{\intercolsep}\hphantom{}}} & 
    \multicolumn{1}{l}{\cellcolor{lightblue}{\bf !Type\hspace{\intercolsep}\hphantom{}}} & 
    \multicolumn{1}{l}{\cellcolor{lightblue}{\bf !Parent\hspace{\intercolsep}\hphantom{}}} & 
    \multicolumn{1}{l}{\cellcolor{lightblue}{\bf !Format\hspace{\intercolsep}\hphantom{}}} & 
    \multicolumn{1}{l|}{\cellcolor{lightblue}{\bf !Verbose name\hspace{\intercolsep}\hphantom{}}} \\
    \midrule
    
    \multicolumn{1}{|l}{\cellcolor{lightergrey}{\bf Gene\hspace{\intercolsep}\hphantom{}}} & 
    \multicolumn{1}{l}{\cellcolor{lightergrey}{\bf Class\hspace{\intercolsep}\hphantom{}}} & 
    \multicolumn{1}{l}{\cellcolor{lightergrey}{\hspace{\intercolsep}\hphantom{}}} & 
    \multicolumn{1}{l}{\cellcolor{lightergrey}{\bf row}} & 
    \multicolumn{1}{l|}{\cellcolor{lightergrey}{\bf Gene\hspace{\intercolsep}\hphantom{}}} \\
    \arrayrulecolor{lightgray}\midrule
    id 
    \hspace{\intercolsep} & Attribute 
    \hspace{\intercolsep} & Gene 
    \hspace{\intercolsep} & String(primary=True, unique=True) 
    \hspace{\intercolsep} & Id 
    \hspace{\intercolsep} \\
    \arrayrulecolor{lightgray}\midrule
    
    symbol 
    \hspace{\intercolsep} & Attribute 
    \hspace{\intercolsep} & Gene 
    \hspace{\intercolsep} & String 
    \hspace{\intercolsep} & Symbol
    \hspace{\intercolsep} \\
    \arrayrulecolor{lightgray}\midrule
    
    location 
    \hspace{\intercolsep} & Attribute 
    \hspace{\intercolsep} & Gene 
    \hspace{\intercolsep} & OneToOne('Location', related\_name='genes') 
    \hspace{\intercolsep} & Location 
    \hspace{\intercolsep} \\
    \arrayrulecolor{black}\midrule

    \multicolumn{1}{|l}{\cellcolor{lightergrey}{\bf Transcript\hspace{\intercolsep}\hphantom{}}} & 
    \multicolumn{1}{l}{\cellcolor{lightergrey}{\bf Class\hspace{\intercolsep}\hphantom{}}} & 
    \multicolumn{1}{l}{\cellcolor{lightergrey}{\hspace{\intercolsep}\hphantom{}}} & 
    \multicolumn{1}{l}{\cellcolor{lightergrey}{\bf row\hspace{\intercolsep}\hphantom{}}} & 
    \multicolumn{1}{l|}{\cellcolor{lightergrey}{\bf Transcript\hspace{\intercolsep}\hphantom{}}} \\
    \arrayrulecolor{lightgray}\midrule
    id 
    \hspace{\intercolsep} & Attribute 
    \hspace{\intercolsep} & Transcript 
    \hspace{\intercolsep} & String(primary=True, unique=True) 
    \hspace{\intercolsep} & Id 
    \hspace{\intercolsep} \\
    \arrayrulecolor{lightgray}\midrule
    
    gene 
    \hspace{\intercolsep} & Attribute 
    \hspace{\intercolsep} & Transcript 
    \hspace{\intercolsep} & ManyToOne('Gene', related\_name='transcripts') 
    \hspace{\intercolsep} & Gene 
    \hspace{\intercolsep} \\
    \arrayrulecolor{lightgray}\midrule
    
    location 
    \hspace{\intercolsep} & Attribute 
    \hspace{\intercolsep} & Transcript 
    \hspace{\intercolsep} & OneToOne('Location', related\_name='transcripts') 
    \hspace{\intercolsep} & Location 
    \hspace{\intercolsep} \\
    \arrayrulecolor{black}\midrule
    
    \multicolumn{1}{|l}{\cellcolor{lightergrey}{\bf Location\hspace{\intercolsep}\hphantom{}}} & 
    \multicolumn{1}{l}{\cellcolor{lightergrey}{\bf Class\hspace{\intercolsep}\hphantom{}}} & 
    \multicolumn{1}{l}{\cellcolor{lightergrey}{\hspace{\intercolsep}\hphantom{}}} & 
    \multicolumn{1}{l}{\cellcolor{lightergrey}{\bf multiple\_cells\hspace{\intercolsep}\hphantom{}}} & 
    \multicolumn{1}{l|}{\cellcolor{lightergrey}{\bf Location\hspace{\intercolsep}\hphantom{}}} \\
    \arrayrulecolor{lightgray}\midrule
    chromosome 
    \hspace{\intercolsep} & Attribute 
    \hspace{\intercolsep} & Location 
    \hspace{\intercolsep} & String 
    \hspace{\intercolsep} & Chromosome 
    \hspace{\intercolsep} \\
    \arrayrulecolor{lightgray}\midrule
    
    five\_prime
    \hspace{\intercolsep} & Attribute 
    \hspace{\intercolsep} & Location 
    \hspace{\intercolsep} & PositiveInteger(primary=True, unique=True) 
    \hspace{\intercolsep} & 5' 
    \hspace{\intercolsep} \\
    \arrayrulecolor{lightgray}\midrule
    
    three\_prime 
    \hspace{\intercolsep} & Attribute 
    \hspace{\intercolsep} & Location 
    \hspace{\intercolsep} & PositiveInteger 
    \hspace{\intercolsep} & 3' 
    \hspace{\intercolsep} \\
    \arrayrulecolor{black}\bottomrule
    \end{tabularx}
    
    \caption{
    \textbf{Overview of how {\ObjTables} helps authors create high-quality spreadsheets and how {\ObjTables} helps other investigators reuse them.}
    This figure uses a dataset of human genes and their splice variants as an example.
    First, authors use programs such as Excel to write their data (e.g., genes and splice variants) to worksheets (\textbf{a}, \textbf{f}, \textbf{g}). Second, authors define an additional schema worksheet that describes the types of objects in their spreadsheets (e.g., genes, splice variants), their attributes (e.g., id, location), and their relationships (variants of each gene; \textbf{c}, \textbf{h}). Third, authors use markup syntax (e.g., `!!Gene`, `!Id`, \textbf{a}, \textbf{f}, \textbf{g}) to indicate the types of data in each worksheet and column. Fourth, authors use the {\ObjTables} software to error check their data (\textbf{d}). Finally, other investigators use the {\ObjTables} software to compare and compose spreadsheets and map spreadsheets to high-level data structures (\textbf{d}, \textbf{i}). 
    }
    \label{fig:fig}
\end{figure}

We developed the {\ObjTables} toolkit (\href{https://objtables.org}{https://\allowbreak{}obj\allowbreak{}tables.\allowbreak{}org}) to both help authors create high-quality spreadsheets and help other researchers reuse them. {\ObjTables} facilitates data quality control and reuse by combining spreadsheets (\hyperref[fig:fig]{Fig.~\ref*{fig:fig}a}) with rigorous schemas (\hyperref[fig:fig]{Fig.~\ref*{fig:fig}c}) that describe types of objects represented by a spreadsheet and software tools for using schemas to systematically error check, compare, and compose data and translate spreadsheets to computational objects suitable for analysis with tools such as Python (\hyperref[fig:fig]{Fig.~\ref*{fig:fig}b,d}).

Using a dataset of two genes and their splice variants as an example, \hyperref[fig:fig]{Fig.~\ref*{fig:fig}} illustrates how {\ObjTables} assists researchers. First, authors use programs such as Excel to write data (e.g., genes and splice variants) to one or more worksheets (\hyperref[fig:fig]{Fig.~\ref*{fig:fig}f,g}). {\ObjTables} supports the Office Open Spreadsheet XML (XLSX) format \cite{ecma376, iso29500}, as well as sets of comma- and tab-separated values (CSV, TSV) files. Second, authors define an additional schema worksheet that describes the types of objects in their worksheets (e.g., genes and splice variants) and their attributes (e.g., id, location) and relationships (variants of each gene; \hyperref[fig:fig]{Fig.~\ref*{fig:fig}h}). {\ObjTables} supports numerous types of attributes for scientific information such as mathematical expressions and chemical equations. Third, authors use markup syntax that begins with exclamation points to indicate the type of object and attribute in each worksheet and column (e.g., `!!Gene', `!Id`; \hyperref[fig:fig]{Fig.~\ref*{fig:fig}f,g}). Fourth, authors use the {\ObjTables} software to error check their data. Finally, other researchers use the {\ObjTables} software to systematically compare and compose spreadsheets, as well as translate spreadsheets to high-level data structures (\hyperref[fig:fig]{Fig.~\ref*{fig:fig}i}). The {\ObjTables} software is available as a web application, command-line program, web service, and Python package.

To make data readable by people and machines, {\ObjTables} supports common spreadsheet layouts such as grouped columns with bi-level headings; grammars for encoding information such as reaction equations into strings in cells; and transposed worksheets that encode records into columns rather than rows. The {\ObjTables} software can also pretty-print spreadsheets by highlighting column headings, embedding descriptions of each column into notes on their headings, and creating a table of contents worksheet (\hyperref[fig:fig]{Fig.~\ref*{fig:fig}e}). To help researchers develop spreadsheets collaboratively, {\ObjTables} also provides tools for tracking versions of schemas and migrating spreadsheets to new versions of their schemas. {\ObjTables} can also export datasets to JSON and YAML.

{\ObjTables} is openly available at \href{https://objtables.org}{https://\allowbreak{}obj\allowbreak{}tables.\allowbreak{}org}. \textcolor{linkcolor}{Section~1 of the Supplementary Information} and the {\ObjTables} website contain examples, a tutorial, and documentation.

Going forward, we aim to make {\ObjTables} more powerful and easier to use by developing attributes for additional data types, supporting additional layouts, developing libraries for additional programming languages, developing a repository for schemas, and developing a graphical user interface (\textcolor{linkcolor}{Section~5 of the Supplementary Information}).

Realizing the full potential of {\ObjTables} as a platform for data reuse will require community adoption. To encourage community participation, {\ObjTables} is an open project, and we invite researchers to try {\ObjTables}, request help via GitHub issues, provide input via pull requests, or join the {\ObjTables} team. Long-term, we aim to push the community to publish reusable spreadsheets by lobbying journals and data repositories to require reusable formats such as {\ObjTables}.

While {\ObjTables} requires additional effort from authors, we believe that the benefits to other investigators of higher-quality and more reusable data are worth the effort. We also believe that authors who use {\ObjTables} will be rewarded with more citations. Due to the popularity of spreadsheets, we believe that {\ObjTables} can increase the reusability of a substantial fraction of new scientific data. Once adoption of {\ObjTables} reaches a critical mass within a field, {\ObjTables} could also enable unprecedented secondary analyses such as meta-analyses of data reported by multiple investigators, comparative analyses of multiple organisms, and multi-dimensional analyses of large systems. For example, we are using {\ObjTables} to merge kinetic and thermodynamic information about \textit{Escherichia coli} into a more predictive genome-scale model of its metabolism (\textcolor{linkcolor}{Section~3.1 of the Supplementary Information}).

By making it easier to develop new formats for new types of data, we believe that {\ObjTables} can also accelerate emerging fields of science. For example, we have used {\ObjTables} to develop WC-Lang, a format for whole-cell models that has become an essential tool for collaboratively developing models (\textcolor{linkcolor}{Section~3.2 of the Supplementary Information}).

\section*{Availability}
{\ObjTables} is freely and openly available under the MIT license. 
The web application is available at \href{https://objtables.org/app}{https://\allowbreak{}obj\allowbreak{}tables.\allowbreak{}org/\allowbreak{}app}. The web service is available at \href{https://objtables.org/api}{https://\allowbreak{}obj\allowbreak{}tables.\allowbreak{}org/\allowbreak{}api}. The command-line program and Python package are available from PyPI at \href{https://pypi.org/project/obj-tables/}{https://\allowbreak{}pypi.\allowbreak{}org/\allowbreak{}pro\allowbreak{}ject/\allowbreak{}obj-\allowbreak{}ta\allowbreak{}bles}. The source code is available at \href{https://github.com/KarrLab/obj_tables}{https://\allowbreak{}git\allowbreak{}hub.\allowbreak{}com/Karr\allowbreak{}Lab/\allowbreak{}obj\_\allowbreak{}ta\allowbreak{}bles}. Examples, tutorials, and documentation are available at \href{https://objtables.org}{https://\allowbreak{}obj\allowbreak{}tables.\allowbreak{}org}. 

\section*{Acknowledgements}
We thank Yin Hoon Chew, Paul Lang, Timo Lubitz, Elad Noor, and the Center for Reproducible Biomedical Modeling for thoughtful feedback. This work was supported by National Institutes of Health grants R35GM119771 and P41EB023912 and National Science Foundation grant 1649014 to J.R.K, German Research Foundation grant Ll 1676/2-2 to W.L, and the Icahn Institute of Data Science and Genomic Technology.



\section*{Author contributions}
J.K., W.L. and A.G. conceived of the project. 
J.K. and W.L. designed the formats. 
J.K., A.G., J.S., and B.S. implemented the software.
J.K. and W.L. developed the case studies. 
J.K. wrote the manuscript.
All of the authors contributed to and approved this manuscript.


\section*{Competing interests}
The authors declare no competing interests.

\section*{Supplementary information}

\textbf{Supplementary information}\\
Descriptions of the {\ObjTables} format for schemas, markup syntax for spreadsheets, attributes, and software for validating, comparing, and composing spreadsheets; case studies of using {\ObjTables} to represent, quality control, and compose metabolomic data and develop a format for whole-cell models; comparison of {\ObjTables} with other tools; and future directions for the development of {\ObjTables}.

\textbf{Supplementary dataset 1}\\
XLSX version of the spreadsheet and schema in \hyperref[fig:fig]{Fig.~\ref*{fig:fig}e--h}.

\textbf{Supplementary dataset 2}\\
CSV version of the spreadsheet and schema in \hyperref[fig:fig]{Fig.~\ref*{fig:fig}e--h}.

\textbf{Supplementary dataset 3}\\
TSV version of the spreadsheet and schema in \hyperref[fig:fig]{Fig.~\ref*{fig:fig}e--h}.

\section*{References}
\renewcommand{\section}[2]{}%

\end{document}